\documentclass[journal,twocolumn]{IEEEtran}
\usepackage{amsmath}
\usepackage{amssymb}
\usepackage{graphicx}

\usepackage[linesnumbered,ruled]{algorithm2e}

\newtheorem{proposition}{Proposition}
\begin{document}
\title{On  Locality of Generalized Reed-Muller Codes over the Broadcast Erasure Channel}

\author{Amira Alloum, Sian-Jheng Lin, and Tareq Y. Al-Naffouri\thanks{Amira Alloum is with Nokia Bell Laboratories, France; Sian-Jheng Lin is with School of Information Science and Technology, University of Science and Technology of China (USTC), China and the Electrical Engineering Department, KAUST; Tareq Y. Al-Naffouri is with the Electrical Engineering Department, King Abdullah University of Science and Technology (KAUST). }}

\IEEEoverridecommandlockouts
\IEEEpubid{\makebox[\columnwidth]{
978-1-4673-9044-6/16\$31.00~
\copyright2016
IEEE \hfill} \hspace{\columnsep}\makebox[\columnwidth]{ }} 

\maketitle

\begin{abstract}
	
	  One to Many communications are expected to be among  the  killer applications for  the currently discussed 5G standard. 
	    The usage of coding mechanisms is impacting broadcasting standard quality,  as coding is involved at several levels of the stack, and more specifically at the application layer where Rateless, LDPC, Reed Slomon codes and network coding schemes have been extensively studied, optimized and standardized in the past. Beyond   reusing, extending or adapting existing  application layer packet coding mechanisms based on previous  schemes  and  designed  for the foregoing LTE or other broadcasting standards; our purpose is to investigate the use of Generalized Reed Muller codes and the value of their locality property in their progressive decoding for Broadcast/Multicast communication schemes with real time video delivery. Our results are meant to bring insight into the use of locally decodable codes in Broadcasting.   
\end{abstract}
%%%%%%%%%%%%%%%%%%%%%%%%%%%%%%%%%%%%%%%%%%%%%%%%%%%%%%%%%%%%%%%%
%%%%%%%%%%%%%%%%%%%%%%%%%%%%%%%%%%%%%%%%%%%%%%%%%%%%%%%%%%%%%%%%
\section{Introduction: Motivation and Related Work}

In  various Broadcasting standard reliable communication is achieved thanks to the use of coding mechanisms  at different levels of the network layers  either for overcoming the channel impairments or recovering lost symbols and packets. %for multicast , broadcast and unicast transmissions. 
At the physical layer incremental redundancy HARQ mechanisms have been thoroughly studied using LDPC codes \cite{soljanin2005incremental}, and also standardized in UMTS and LTE using  Turbo codes  \cite{spec3GPP}.\\
At upper layers historical Reed-Solomon~(RS) codes have been first used in one to many communication standards as DVB-SH thanks to their MDS property . After the advent of LT and Raptor codes \cite{shokrollahi2006raptor} , rateless codes have been widely adopted as a reference for upper layer forward error correction~(FEC) and specified for many Broadcast standards like DVB-SH, LTE-EMBMS. Their success is technically motivated by their universality as they are optimal over all erasure channels discarding any constraint of feedback . Another line of work proposed the use of Network Coding for packet loss concealment in Broadcast applications, where  systematic binary network coding has been proved to be the best tradeoff in terms of  complexity, throughput and decoding delay compared to classical straightforward network coding in finite fields\cite{jones2015binary}.
The interest of combining multilayer coding has been analyzed in \cite{heindlmaier2014isn} and is revealed to be more beneficial for the multicast scenario than the point to point communication usecase.

Besides some broadcasting standards like MBMS specify  a uni-cast repair mechanism after a limited broadcast delivery phase. During this phase, coding can be used on the top of TCP protocol for uni-cast communication where network codes or rateless codes  can be used  on top of physical layer in a separate or a cross layer approach \cite{sundararajan2009network}.

In the context of recovering the packets or symbols lost during bad channel realizations, the common channel model considered to evaluate the recovery performance is the symbol~(bit or block) erasure channel. On another side, most proofs derived for demonstrating capacity achieving properties have been derived over the erasure channel first before their extensions to other Binary Memoryless Channels (BMC). 

 While the common belief that random constructions are the unique road for capacity and that deterministic structures might be unable to honor the bet. This common belief has been abolished with the discovery of polar codes which are  based on Reed Muller codes constructions.
%The race to the capacity, has started with the discovery of turbo codes \cite{berrou2003turbo} and the renaissance of Gallager's LDPC codes\cite{richardson2003renaissance} where their capacity achieving property using iterative decoding has been proven and investigated for more than a decade \cite{richardson2001capacity} cite{richardson2001design}~\cite{chung2001design}  \cite{boutros2006quasi} while the common belief that random constructions are the unique road for capacity and that deterministic structures might be unable to honor the bet. This common belief has been abolished with the discovery of polar codes which are the first probably capacity achieving codes \cite{arikan2009channel},which are  derivatives of Reed Muller codes.%
  Consequently the invention of Polar codes rekindled the flame for algebraic codes among coding theory community, especially for Reed Muller Codes that have been introduced 50 years ago by Reed and Muller \cite{Reed1954} \cite{muller1953metric}, then  have been generalized by Kasami and Lin in~\cite{kasami1968new}. This class of code is special for having kept   in the meanwhile the common interest of theoretical computer science and information theory communities concurrently. 
More recently Reed Muller codes have been proved to be capacity achieving over the binary and block erasure channel~\cite{kumar2015reed} \cite{kudekar2015reed}.

Reed Muller codes are appreciated for being good practical extended cyclic codes meeting BCH codes in some instances of their generalization; they exhibit good geometrical and nesting properties and are good basis for constructing other codes  , as they are a part of the LTE standard with the encoding of channel quality control informations \cite{spec3GPP} and have been used for Power Control issues for OFDM .
Besides, Reed Muller codes exhibit an additional property called Locality which has been leveraged recently for coded based unconditionally secure protocols considered in theoretical computer science and cryptography communities~\cite{yekhanin2011locally}.   
Locality feature consist in the ability of retrieving a particular symbol of a coded message by looking only at $\ell <K$ positions of its encoding, where $\ell$ is known as the locality parameter and $k$ denotes the dimension of the code.Besides,  Locally decodable codes have been mentionned in \cite{lott2007hybrid} as a potential candidate for improving the power budget of HARQ schemes.

The recent results cited above, led us to rise some questions regarding the practical use of Reed Muller Codes in upper layer coding mechanisms and more specifically questioning about the value of  locality in this context.\\

Our focus is on broadcasting for streaming services, where are required low complexity decoding algorithms in order to enable reception for energy constrained devices, together with  short decoding delays in order to trigger the content reception as soon as possible without sacrificing throughput.In previous studies the decoding delay reduction is reached thanks to the use of systematic encoding constructions with progressive decoding . The complexity is decreased by the use of binary fields for network coding \cite{jones2015binary}.

In our scheme we  consider the decoding and delay performance of Generalized Reed Muller (~GRM) codes over the block erasure channel under a locally symbol wise decoding algorithm. This is motivated by the fact that the GRM codes can be systematically encoded, and enable a progressive decoding thanks to the locality property earlier than word-wise Reed Solomon algebraic decoding.  

The main purpose of our analysis, is to investigate the costs and the benefits of considering the locality in packets recovery using Generalized Reed Muller codes. 

The paper is organized as follows: In section \ref{sec:Intro} we introduce the model and notations related to the Generalized Reed Muller Codes, Maximum likelihood (~ML) decoder is described in section \ref{sec:gaussian}, the exhaustive local decoder (~LD) is derived and described in section \ref{sec:localdec} , complexities of both ML and LD schemes are evaluated  in section \ref{sec:complexity}. We show simulations results  and discuss the decoding performances  in section \ref{sec:simulations} and conclude in  section \ref{sec:ccl}.
\vspace{-0.3cm}
%%%%%%%%%%%%%%%%%%%%%%%%%%%%%%%%%%%%%%%%%%%%%%%%%%%%%%%%%%%%%%%%
%%%%%%%%%%%%%%%%%%%%%%%%%%%%%%%%%%%%%%%%%%%%%%%%%%%%%%%%%%%%%%%%
\section{Models, Notations and Problem Formulation}
\label{sec:Intro}
%%%%%%%%%%%%%%%%%%%%%%%%%%%%%%%%%%%%%%%%%%%%%%%%%%%%%%%%%%%%%%%%
\subsection{Transmission and Channel Model}
We consider a broadcasting communication standard where the sender is a basestation or  a satellite transmitting $k$ data packets encoded to $k$ systematic packets and $(n-k)$ parity packets using an error correcting code. 

  The receivers are  devices either within a satellite network or wireless cellular network, where each packet is considered by the upper layers either completely received or completely lost based on a CRC (Cyclic Redundancy Check) mechanism.
  
  Accordingly the described practical scenario is modelled as a transmission occurring through a virtual block erasure channel characterised by an erasure  probability $\epsilon$ over  coded symbols in  the  considered finite field  $\mathbf{F}_q$. Consequently each erased packet is related to   an erased coded symbol of $\mathbf{F}_{q}$ within the model.  
      %Let $\epsilon$  denotes the  channel erasure probability.
%%%%%%%%%%%%%%%%%%%%%%%%%%%%%%%%%%%%%%%%%%%%%%%%%%%%%%%%%%%%%%%%
\vspace{-0.3cm}
\subsection{Generalized Reed Muller Codes}
Generalized Reed Muller code are constructed by complete evaluation of low degree multivariate polynomial over a finite field. The code is specified  by parameters $(r,m,q) $ where:
\begin{itemize}
\item $q$ is the alphabet size and is a prime power.
\item $m$ denotes the number of variables.
\item $r\leq q-2$ is the degree of the polynomial.
\end{itemize}
Let $\mathbf{F}_q = \{\gamma_i\}_{i=0}^{q-1}$ denotes a finite field  with  $q$ elements. 
Let us consider the variable vector  $ \mathbf{X}=(X_1,\dots ,X_m) $ where $F(\mathbf{X})\in \mathbf{F}_q[\mathbf{X}]$ denotes an $m$-variable polynomial of degree at most $r\leq q-2$. 
The number of coefficients of $F(\mathbf{X})$ is  the dimension of the code $k={m+r\choose r}$ where the minimum distance is $d=(q-r)q^{m-1}$ and $n=q^m$ is the code length of the code\cite{lachaud1986projective}. 

The $(r,m)$ Reed-Muller code over $\mathbf{F}_q$ is defined as:
\[
(F(\mathbf{P}_1),F(\mathbf{P}_2),\dots ,F(\mathbf{P}_n)),
\]
where each point $\mathbf{P}_i\in \mathbf{F}_q^m$ belongs to the $m$-dimensional affine space over $\mathbf{F}_q$.
Let us consider the vector:
\begin{equation}\label{eq:V}
V=[v_0\dots v_{m-1}]^T\in \mathbf{F}_q^m\setminus \{0\},
\end{equation}
and the point:
\begin{equation}\label{eq:P}
P=[p_0\dots p_{m-1}]^T\in \mathbf{F}_q^m.
\end{equation}
The $q$ elements:
\begin{equation}\label{eq:yi}
y_i=F(P+\gamma_i\cdot V)=F_{P, V}(\gamma_i),\qquad \gamma_i\in \mathbf{F}_q,
\end{equation}
forms a $(q,r+1)$ Reed-Solomon~(RS) codeword of dimension $r+1$ and code length $q$, and $F_{P, V}$ is a univariate polynomial of degree at most $r$. 

Local decoding is possible over a block erasure channel, when $r+1$ received symbols are the evaluations of points aligned on a such a line, then the RS decoding algorithm can occur on the received symbol and perfectly reconstruct the erased $q-(r+1)$ symbols via polynomial interpolations.
%%%%%%%%%%%%%%%%%%%%%%%%%%%%%%%%%%%%%%%%%%%%%%%%%%%%%%%%%%%%%%%%
%%%%%%%%%%%%%%%%%%%%%%%%%%%%%%%%%%%%%%%%%%%%%%%%%%%%%%%%%%%%%%%%
\section{Maximum Likelihood Decoding for Generalized Reed Muller Codes}
\label{sec:gaussian}
The Maximum likelihood decoder over erasure channel fails to recover a given erasure pattern if this pattern contains the support of at least one non zero codewords. Their decoding  consist in  a Gaussian Elimination algorithm over the subsequent linear system.

   The generator matrix $G$ of GRM code is of size $k$-by-$n$, where the $i$-line can be determined with the RM encoder. Let $I_i=[0\dots 0\; 1\; 0\dots 0]$ denote a $k$-element zero vector with $1$ at $i$-th position. By applying the RM encoding approach on $I_i$, the codeword vector, denoted as $R_i$, is the $i$-th row of $G$. If the code is systematic, the generator matrix is in the form $G = \begin{bmatrix} I_k | \mathcal{P} \end{bmatrix}$,
and its corresponding parity-check matrix is written as $H = \begin{bmatrix} -\mathcal{P}^{\top} | I_{n-k} \end{bmatrix}$. 

Given  a sent codeword $C$ from the GRM codebook and a received codeword $Y$ at the output of a block erasure channel , we consider the syndrome constraint  $HY^{\top} = 0$ from which we derive the following linear system  :
\begin{equation}\label{eq:HH}
H_0 {Y_0}^{\top} = D^{\top},
\end{equation}
where ${Y}_0$ denotes the erasure symbols, and $H_0$ consists of the subsequent  columns corresponding to $Y_0$. Our objective is to solve $\bar{Y}_0$ in \eqref{eq:HH}. 

To solve \eqref{eq:HH}, we apply Gaussian elimination on $\begin{bmatrix} H_0| D^{\top}\end{bmatrix}$, to obtain a matrix $\mathcal{Q}$ in the reduced row echelon form. For each row of $\mathcal{Q}$, if the row is in the form $[I_i|d_i]$, then the value of the $i$-th symbol is $d_i$. Gaussian elimination utilizes all parity equations in RM codes, so it performs the optimal performance for erasure channels. However, the algorithm is very slow and thus it is untractable for long codes in finite fields of high orders. 

%%%%%%%%%%%%%%%%%%%%%%%%%%%%%%%%%%%%%%%%%%%%%%%%%%%%%%%%%%%%%%%%
%%%%%%%%%%%%%%%%%%%%%%%%%%%%%%%%%%%%%%%%%%%%%%%%%%%%%%%%%%%%%%%%

\section{Local Decoding  For Generalized Reed Muller Codes}
\label{sec:localdec}

In the present section we derive an exhaustive  local decoding algorithm  for GRM codes.

Let us Consider the set $\Phi$ of received symbols in $\mathbf{F}_{q}$
evaluating the set of points $\Phi'$ in $\mathbf{F}_{q}^m$. Our algorithm consists in exhaustively and sequentially  search in the space, all the lines including $r+1$ received symbols evaluating $r+1$ aligned points of $\Phi'$ in order to  apply Reed Solomon interpolation algorithm. %and  recover the erasures thanks to the fact that  each line includes the support of a Reed solomon codeword. 
For thus,  it is required to enumerate all $r+1$ received symbols defining a support for a $(q,r+1)$ Reed Solomon sub-code among all the RS subcodes nested  in a Reed Muller codeword.

In order to perform the exhaustive enumeration of the lines in the space, we build a partition of the space similar to the one used for the construction of Projective Reed Muller codes  \cite{lachaud1986projective}. 
Our approach paves the way to the following proposition:

\begin{proposition}
	The support of Generalized Reed Muller codeword of codelength $n=q^m$ includes  the support of $q^{m-1}\times \frac{q^m-1}{q-1}$  number of $(q,r+1)$ Reed Solomon subcodes. 
\end{proposition}

\textbf{Proof}: \\
		In order to enumeratethe lines $\mathbf{P}+\gamma_i\cdot \mathbf{V}$, one can consider $q^m-1$ instances of $\mathbf{V}$ and $q^m$ instances of $\mathbf{P}$ where the affine space would include $(q^m-1)\times q^m$ lines. Though, this is not a tight bound as some lines are counted multiple times.

	Let us consider the partition of the affine space $\mathbf{F}_{q}^m$  where $\mathbf{V}_0\cup \dots \cup \mathbf{V}_{m-1}=\mathbf{F}_q^m\setminus \{0\}$ and defined as :
	
\begin{eqnarray*}
\begin{aligned}
\mathbf{V}_i=\{ V_i=[0\dots 0\; 1\; v_{i+1}\dots v_{m-1}]^T| v_j \in \mathbf{F}_q, \\
 i \leq   j\leq m-1, v_i\neq 0\ \}
\end{aligned}
\end{eqnarray*}

where $\mathbf{V}_i$ denotes a set of $m$-element vectors where the first $i$ elements are zeros, and $v_i\neq 0$. The number of instances per subspace is$|\mathbf{V}_j|=q^{m-j-1}$. Accordingly, the number of possible vectors $V$ is  $q^{m-1}+q^{m-2}+\dots +1=\frac{q^m-1}{q-1}$.		
	
	Let us consider the partition of $P\in \mathbf{F}_{q}^m$. The set of evaluation points is:
		\[
		P+\gamma_i\cdot V_j=P-p_iV_j+(\gamma_i+p_i)\cdot V_j=P_i'+\gamma_i'\cdot V_j
		\]
		where  $ P_i'=P-p_iV_j=[p_0'\dots p_{m-1}']^T$ is a vector with $i-th$ element $p_i'=0$.
	    the set of values of $P_i'$ is: 
		 \begin{equation*}\begin{aligned}
		\mathbf{P}_i=\{[p_0\dots p_{m-1}]^T|p_j\in \mathbf{F}_q,0\leq j\leq m-1, j\neq i, p_i=0\},
		 \end{aligned}
		\end{equation*}
		and $|\mathbf{P}_i|=q^{m-1}$	.
		
		Consequently we have enumerated $q^{m-1}\times \frac{q^m-1}{q-1}$ different lines in the affine space $\mathbf{F}_{q}^m$and so as much Reed Solomon subcodes nested in the space of  an RM codeword. $\square$

Notably, when $r=q-2$, the local decoding considers a $(q,q-1)$ RS codeword of one parity symbol. In this case, the set of received symbols of \eqref{eq:yi} satisfy the parity  equation
\[
F_{P, V}(\gamma_0)+\dots +F_{P, V}(\gamma_{q-1})=0.
\]
Thus, we can decode one symbol via only field additions, and the field multiplications are unnecessary.
Based on the previous discussion, the sequential exhaustive local decoding algorithm is described in Algorithm \ref{alg:Enc}.
\begin{algorithm}[h]
\caption{\label{alg:Enc} Decoding Reed-Muller codes via locally decoding approach}
\KwData{A set of RM symbols $\{(e_i, \mathbf{U}_i)\}_{i=0}^{\ell -1}$, where $e_i=F(\mathbf{U}_i)$ is the symbol and $\mathbf{U}_i\in \mathbf{F}_q^{m}$ indicates its evaluation point}
\KwResult{A codeword $M'$ via locally decoding approach}
$z=1$\\
\While{$z=1$}{
$z=0$\\
\For{$i=0$ \KwTo $(m-1)$}{
\For{$V\in \mathbf{F}_q^{i-1}$}{
$\mathbf{V}=[\underbrace{0\dots 0}_{m-i}\; 1\; V]$\\
\For{$P_0\in \mathbf{F}_q^{m-i-2}$}{
\For{$P_1\in \mathbf{F}_q^{i-1}$}{
$\mathbf{P}=[P_0\; 0\; P_1]$\\
Count the number of received symbols $R$ on $(\mathbf{P}+\gamma_i\cdot \mathbf{V})$, for $\gamma_i\in \mathbf{F}_q$.\\
\If{$R\geq r+1$}{
Apply $(q, r+1)$ RS decoding to interpolate the lost symbols.\\
$z=1$
}
}
}
}
}
}
\Return M'
\end{algorithm}
%%%%%%%%%%%%%%%%%%%%%%%%%%%%%%%%%%%%%%%%%%%%%%%%%%%%%%%%%%%%%%%%
%%%%%%%%%%%%%%%%%%%%%%%%%%%%%%%%%%%%%%%%%%%%%%%%%%%%%%%%%%%%%%%%
%\section{Theoretical Perfomance Analysis}
%%%%%%%%%%%%%%%%%%%%%%%%%%%%%%%%%%%%%%%%%%%%%%%%%%%%%%%%%%%%%%%%
%%%%%%%%%%%%%%%%%%%%%%%%%%%%%%%%%%%%%%%%%%%%%%%%%%%%%%%%%%%%%%%%
\section{Complexity Analysis }
\label{sec:complexity}
This section discusses the computational complexities of the local decoding algorithm~(LD) and  the Gaussian elimination~(GE) algorithm described in the previous sections. 

     The maximum likelihood decoder is known to be the one of the GE  applied on the matrix $\begin{bmatrix} H_0| D^{\top}\end{bmatrix}$, which requires $O(n^3)$ operations. However regarding the  proposed LD decoder the theoretical complexity is  not straightforward, as the number of loops (line 1-19) is not analytically quantified so far.
    
   Nevertheless, we introduce an alternative related decoding strategy, termed progressive local decoding~(PLD), that triggers the decoding  on the fly with the reception   of a fraction of the codeword on a symbol by symbol basis . 
	%Unlike the traditional approach that will start the decoding algorithm after receiving the whole codeword, the progressive approach triggers the decoding  on the fly with the reception   of a fraction of the codeword on a symbol by symbol basis.
	
Once a symbol is received, PLD checks all lines across this symbol. Once a line includes  $r+1$ either received or recovered symbols , RS decoding  is applied. 
For a codeword of $(r,m,q)$ RM code, there are $q^{m-1}$ lines across a symbol. With the use of fast fourier transform (FFTs) techniques, RS erasure decoding requires $O(q\lg (q))$ operations ~\cite{LinFFT}. Hence, the global per symbol recovery operation  requires $O(q^m\lg (q))$ operations in each round. As the decoder will receive at most $n$ symbols, the complexity is quasi quadratic and no more than $O(nq^m\lg (q))=O(n^2\lg (q))$. Consequently the PLD achieves lower complexity than GE does by around one order. 
\begin{figure}[h]
\vspace{-0.5cm}
\begin{center}
\includegraphics[width=0.7\columnwidth]{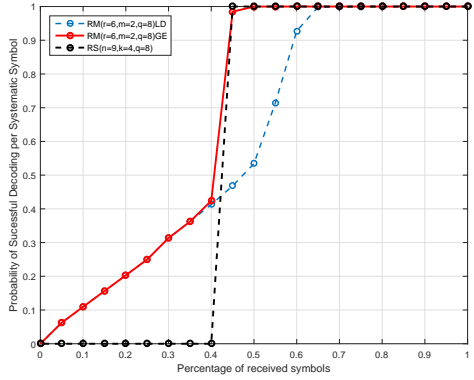}
\caption{\label{sucessprobfig}Probability of Sucessful Decoding Vs percentage of received symbols.}
\vspace{-0.7cm}
\end{center}
\end{figure}

%%%%%%%%%%%%%%%%%%%%%%%%%%%%%%%%%%%%%%%%%%%%%%%%%%%%%%%%%%%%%%%%
%%%%%%%%%%%%%%%%%%%%%%%%%%%%%%%%%%%%%%%%%%%%%%%%%%%%%%%%%%%%%%%%
\section{Simulation Results and Discussion}
\label{sec:simulations}
\vspace{-0.2cm}
In the present section we show simulation results assessing the local decoding (~LD)  performance of  systematic GRM codes over the block erasure channel. 
%For thus, we compare the (~LD) exhaustive decoder to the GE decoder for systematic GRM constructions,  and also with Reed Solomon interpolation  decoding. 

We have selected metrics emphasizing the potential of the  local decoder for progressive packet recovery  in streaming applications \cite{jones2015binary}. For instance the  metric represented in  Figure \ref{sucessprobfig} is the probability of successful decoding per information symbol varying with  percentage of received symbols.
In this scheme we consider LD and GE decoding of the  $(r=6,m=2,q=8)$ RM code compared with interpolation decoding of the  $(n=9,k=4, q=8)$ RS code in $\mathbf{F}_8$. We chose same fields orders to aline  erasure dimension over  the channel.\\
The results in Figure \ref{sucessprobfig}  show that  GE and local decoding execute partial decoding  before the reception of $k$ symbols and before building a full rank linear system for the GE. The locality being $r+1=7$ symbols, the local decoder triggers recovery when receiving $10$ percent of transmitted symbols which is  lower than the code dimension  that is  $43,37$ percent of the transmission as $k=28$. 

%As per example in the region  $]0: 0,39]$ of received symbols
%Accordingly the decoder start decoding at the reception of a number of symbols lower than the dimension of the code $k=28$.
It is also exhibited that LD under-performs  the RS decoder when reception is beyond the code dimension. However GE performs as good as the RS decoder and better than LD  decoding after building a full rank linear system, an erasure point that we refer to as the full rank threshold $\epsilon^{*}$.

\begin{figure}[h]
\vspace{-0.5cm}
\begin{center}
%[width=0.3\columnwidth,angle=270 ]
\includegraphics [width=0.7\columnwidth]{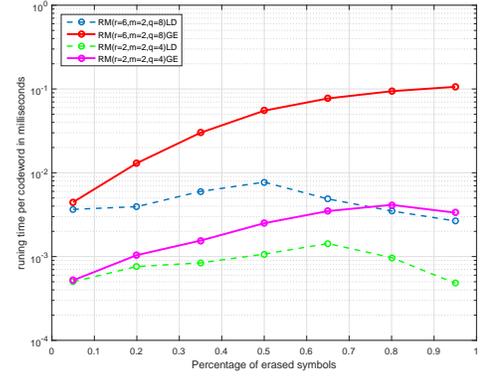}
\caption{\label{complexityfig}Runing time for RM(r=6,m=2, q=8), RM(r=2,m=2, q=4).}
\vspace{-0.5cm}
\end{center}
\end{figure}

Figure \ref{complexityfig}  shows  the  computational cost in terms of time per codeword varying with the percentage of erased symbols. In line with the theoretical analysis we find out that the GE decoding requires more operations than the LD and that the complexity gap is  getting increased with the field order .

Accordingly we conclude that LD is an alternative to the GE decoder  for systematic locallly decodable codes before the full rank threshold $\epsilon^{*}$.The LD enhances the information recovery delay at lower complexity than  ML decoder and  with equal performance. However beyond $\epsilon^{*}$ the system should switch to the GE to get the best performance in a combined architecture LD-GE .
 A further comparison  to systematic random network coding is worth to be investigated. Whereas GRM codes have the inherent benefit of being systematic, and locally decodable with a mitigated complexity.

\vspace{-0.3cm}
%%%%%%%%%%%%%%%%%%%%%%%%%%%%%%%%%%%%%%%%%%%%%%%%%%%%%%%%%%%%%%%%
%%%%%%%%%%%%%%%%%%%%%%%%%%%%%%%%%%%%%%%%%%%%%%%%%%%%%%%%%%%%%%%%
\section{Conclusions and Further Work}
\label{sec:ccl}
%Rate Issue.
%Open avenues for other LDC constructions.%Analytical Analysis
In this paper, we  evaluated  the value of locality property in  GRM codes for packet recovery in Broadcast Applications.
We  revealed that locality and systematic encoding of GRM codes are valuable assets for progressive decoding in streaming usecases. 
However, among the  drawbacks we should mention about  GRM  codes is their decreasing rate when code-length get longer. Therefore investigating other locally decodable constructions  at higher rates may be an interesting future  research avenue for broadcasting applications.
\vspace{-0.2cm}
% either  based on Reed Muller codes or not as  expander codes  
%We derived an exhaustive local decoding based on decomposition of Reed Muller codes as partition of Reed Solomon sub codes.
%We demonstrated that the local decoder performs as good as the GE decoding before the full rank threshold at a lower %complexity in  a region where Reed Solomon fails to recover any packet. Accordingly a combined LD-GE decoder seems to be %the best tradeoff  in terms of delay of systematic packet recovery, decoding performance and complexity.    

%The probability of successful decoding in terms of percentage of received symbols 
%is evaluated as  the rate of recovered symbols over erased symbols varying with the percentage of received symbols for a $(r=6,m=2)$ Reed-Muller code in $\mathbf{F}_8$ compared with $(n=9,k=4)$ RS code in $\mathbf{F}_8$.
%
%We observe that decoding is possible in the range $[0.25: 0,5625[$ , starting from the reception of $16$ symbols while the locality is $r+1=7$. Accordingly the decoder start decoding at the reception of a number of symbols lower than the dimension of the code $k=28$.
%
%However as the decoder is symbol wise and not Maximum Distance Separable as the Reed Solomon code, it underperforms the Reed Solomon decoding algorithm when the number of received symbols is greater than the dimension of the code and this holds for the proposed exhaustive symbol wise decoding algorithm as whether we consider a MAP decoding for erasure channel based on gaussian elimination.
%\bibliographystyle{plain}

\bibliographystyle{IEEEtran}
\bibliography{IEEEabrv,refs}
\end{document}